%% file: artigo-igor.tex
\documentclass{sbrt2018eng}
\include{usepackages}

\begin{document}

\title{C-RAN Virtualization with OpenAirInterface}

\author{Igor Trindade, Cleverson Nahum, Camila Novaes, Daniel Cederholm, Gyanesh Patra and Aldebaro Klautau 
	\thanks{Igor Trindade, Cleverson Nahum, Camila Novaes and Aldebaro Klautau are with 
		LASSE - Telecommunications, Automation and Electronics Research and Development Center, Bel\'em-PA, Brazil, E-mails: \{igor.trindade,camila.novaes.silva\}@itec.ufpa.br, and \{aldebaro, cleversonahum\}@ufpa.br. %This work was partially supported by CNPq (XX/XXXXX-X).} 
}}

\maketitle

\markboth{XXXVII SIMP\'OSIO BRASILEIRO DE TELECOMUNICA\c{C}\~OES E PROCESSAMENTO DE SINAIS - SBrT2019, 29/09/2019--02/10/2019, PETR\'OPOLIS, RJ}{XXXVII SIMP\'OSIO BRASILEIRO DE TELECOMUNICA\c{C}\~OES E PROCESSAMENTO DE SINAIS - SBrT2019, 29/09/2019--02/10/2019, PETR\'OPOLIS, RJ}

\begin{abstract}
C-RAN virtualization is a research topic with great interest since it allows to share baseband processing resources. Therefore, in this work, we report the implementation of a virtualized LTE testbed environment of C-RAN by integrating the OpenAirInterface (OAI) with Docker. Using the testbed, we conducted a workload study to understand the computation resource demand of C-RAN software. Virtualization in containers has proven to be effective in creating a functional 4G network which achieves realistic results to facilitate research.
        
\end{abstract}

\begin{keywords}
C-RAN, OpenAirInterface, virtualization.
\end{keywords}

\section{Introduction}

Currently, there is a significant increase in data consumption as a result of an increasing number of mobile devices. The current costs of building and operating a new infrastructure capable of providing the required data rates are higher than the revenue growth rate~\cite{dahlmanet2011lte}.

As a consequence, the concept of Cloud Radio Access Network (C-RAN) is faced as a cost-efficient approach, which aims to implement the mobile networks in a centralized way. Through cloud technologies, such as virtualization and elastic resource management, it is expected that the operators can reduce the network deployment and operating costs, as well as simplify deployment and management of increasingly heterogeneous radio access networks (RANs)~\cite{luo2018cran}. To further reduce costs,  the baseband processing resources in the cloud center could be shared among several antennas.

Therefore, to evaluate C-RAN virtualization, we used OpenAirInterface (OAI), an open source LTE software developed by EURECOM ~\cite{yeoh2016oai}. OAI offers modules that emulate the User Equipment (UE), eNodeB and Evolved Packet Core (EPC). OAI can be used to implement a network by itself or can be used together with commercial modules~\cite{nahum2017emulation}. This paper describes a C-RAN test network created using the OAI platform together with the virtualization technology in Docker containers, evidencing its efficiency in the deployment.

\section{Architecture of OAI C-RAN}

OAI is an open-source software-defined implementation of Evolved Packet Core (EPC) and Evolved Base Station (eNodeB). EPC is the core network, and it includes a Serving Gateway (S-GW), a Packet Data Network Gateway (P-GW), a Mobility Management Entity (MME) and a Home Subscriber Server (HSS). S-GW and P-GW are responsible for routing user data traffic, while MME and HSS are responsible for controlling signaling messages. eNodeB handles the communication between the UE (User Equipment) and EPC. It is consists of two parts: the distributed RRHs (Remote Radio Heads) more antennas deployed at the remote site and the centralized BBU (Base Band Unit) pool hosted in a cloud data center.

\section{Virtualization of OAI}

Virtualization is a process that allows abstraction of computing resources, so the total capacity of a physical machine can be shared among many users, environments or applications. Through virtualization with flexible resource management, network deployment and operation costs can be reduced, as well as it can simplify deployment and management of RANs~\cite{checko2015cranclassic}.

Virtualization in Containers, which can also be defined as containerization~\cite{docker}, is an operating system level virtualization method for deploying and running distributed applications without starting an entire virtual machine (VM) for each use. A container is a lightweight operating system running inside the host system, running instructions native to the core CPU, eliminating the need for instruction level emulation or just in time compilation. Containers provide savings in resource consumption without the overhead of virtualization, while also providing isolation~\cite{dua2014virtualization}.

Moreover, the operation tasks are reduced with the use of containers, there is less effort to maintain environments for the application. Creating an environment is automated. Therefore, fewer manual actions are required. This reduces the risks and increases reliability. Manually keeping a consistent environment across multiple servers is prone to errors and can be a trouble. With containers, it's easy to create multiple instances of an environment because just run the image on the servers. In this way, it is easy to add more nodes to a cluster and scale them horizontally.

\section{Implementation of OAI}
The implementation of OpenAirInterface modules in Docker containers was based on the OAI tutorial in ~\cite{oai}. Two machines were used, which run with Linux Ubuntu LTS 14.04. The kernel 3.19.0-61-low-latency package is also installed to reduce the packets processing latency. Many configurations have been made to allow the execution of OAI RAN, such as disable any power management features in the BIOS, and CPU frequency scaling, such that OAI can operate at maximum 100\% CPU clock in performance mode and avoid staggering.

To deploy all OAI modules in Docker containers, Dockerfiles were created with the packages and builds needed to compile and run the application. Two Docker images were designed to run OAI, one for the RRH and another for the BBU. For EPC, all three modules mentioned in section II were compiled and run in a virtual machine.

It is possible to automate the testbed creation of RRH and BBU containers using Docker compose~\cite{kumar2017docker-compose}, a tool that defines and runs Docker applications with multiple containers. Also, it is possible to define the behavior that Docker will have if one of the containers fails, so in case the RRH or BBU container fails, the Docker will be able to upload another immediately.

For the execution of the containers, it is necessary to give them extended privileges. By default, Docker containers are unprivileged and cannot access any device external device, but a privileged container is provided access to all ~\cite{chang2018hostmode}.

Besides, you need to configure the network of the containers, the Docker network subsystem is pluggable, using drivers. Several drivers exist by default and provide core networking functionality. We use the host driver because when given the option network host at container launch, Docker does not place the container into a separate network namespace; it, therefore, provides the container with full access to the host's network stack (enabling network sniffing, reconfiguration, and so on)~\cite{combe2016privilege}. 

Figure 1 shows the implemented C-RAN testbed scenario. In testbed, the RRH receives a signal from the UE via USRP B210 board connected to USB 3.0 on the physical machine and sends to the BBU pool. 
\begin{figure}[!htb]
	\begin{center}
		\includegraphics[width=1\columnwidth]{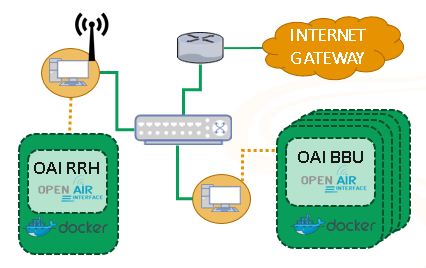}%
		\caption{testbed scenario C-RAN}
		\label{fig:process}
	\end{center}
\end{figure}

\section{Results}

OpenAirInterface was used to create an LTE network, and two containers were built on different machines, which represent the RRH and the BBU. USRP Radio was used to communicate with a smartphone ready for use. Figure 2 shows The fronthaul rate varies according to the UE rate.
\begin{figure}[!htb]
	\begin{center}
		\includegraphics[width=1\columnwidth]{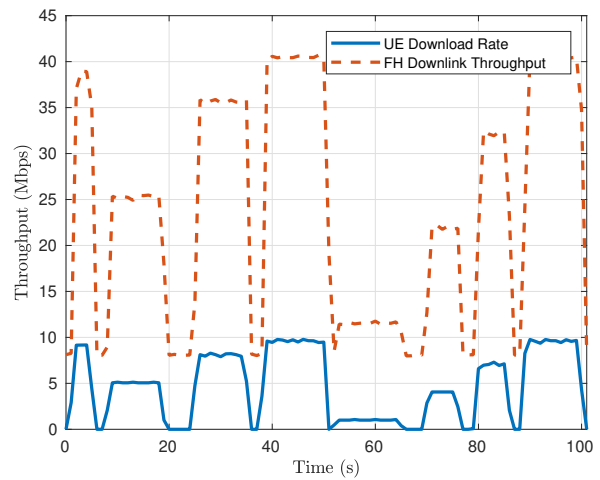}%
		\caption{Variation of fronthaul rate and UE rate.}
			\label{fig:process}
	\end{center}
\end{figure}

The results of the above figure were produced according to the access to videos on YouTube by the smartphone. Thus, we had an easy deployment of OAI LTE components with better use of resources using containers, in addition to generating a flexible testbed that could it easier to started to make researches. Finally, it is possible to test new features on 4G/5G experimental networks in a way fast and efficient. 

\section{Conclusions}
This work presented an automated creation of a virtualized C-RAN test environment using OpenAirInterface. The work aims to provide an automated way to configure an RRH and BBU containers which connects with a commercial UE. Our situation was validated by the use of the smartphone connected to the network generating traffic in fronthaul when accessing videos on YouTube. Finally, it has some advantages over virtual machines, especially about the ability to consume real computing resources to implement the network.

\bibliographystyle{IEEEtran}

\bibliography{references}

\end{document}

%% file: usepackages.tex
\ifdefined\usePortuguese
\usepackage[brazil]{babel}   % Para hifenar em portugus
\usepackage[latin1]{inputenc}% Para poder digitar os acentos da maneira usual:
\else
\fi

\ifdefined\appendices
\else
\ifdefined\addAppendixAsPreamble %Avoid A, B as appendix names and impose Appendix A, Appendix B
\usepackage[titletoc,title]{appendix} %provokes clash with usepackages
\else
\usepackage{appendix} %More support for appendices
\fi
\fi

\usepackage{makeidx}  %to generate indices, I guess
\usepackage{graphicx}
\usepackage{color} %http://en.wikibooks.org/wiki/LaTeX/Colors
\definecolor{darkgreen}{rgb}{0,0.35,0}

\ifdefined\keywords
\else
	\def\keywords{\vspace{.5em}{\bfseries\textit{Index Terms}---\,\relax%
	}}
	\def\endkeywords{\par}
\fi

\usepackage{url} %LIMITED: does not have the text to be displayed as second parameter

\ifdefined\RCSfile %check if Elsevier class
\else
\usepackage{cite} %to be smart when citing multiple references, example: [1-7] instead of [1,2,..]
\fi

\usepackage{multirow} %tables with multiple rows
\usepackage[bbgreekl]{mathbbol}

\ifdefined\amsmath
\else
\usepackage{amsmath}
\fi

\ifdefined\amsfonts
\else
\usepackage{amsfonts}
\fi

\usepackage{amssymb}

\ifdefined\proof
\else
\usepackage{amsthm}
\fi

\ifdefined
\else
\ifdefined\akbook
\ifdefined\lulu
\typearea[6mm]{1} % 6mm for spine
\else
\usepackage[a4paper, top=3cm, bottom=3.4cm, left=1.8cm, right=2.5cm]{geometry}
\fi
\fi
\fi

\usepackage{mathrsfs} %math alphabet I will use for sets
\usepackage{makeidx}  %to generate indices, I guess
\usepackage{graphicx,color}

\usepackage{multirow} %tables with multiple rows
\usepackage{listings} % to list source code: http://www.usq.edu.au/users/leis/notes/latex/code.html
\usepackage{ifpdf} %The package provides the switch \ifpdf:
\ifpdf

\ifdefined\hyperref
\else
\usepackage[pdftex,colorlinks]{hyperref}
\fi

\usepackage{epstopdf}
\hypersetup{%
pdftitle={Some title},
pdfauthor={Aldebaro - LASSE - UFPA},
pdfkeywords={DSP,Signal},
pdfstartview={FitH}, %% <--
urlcolor=black,
linkcolor=black,
citecolor=black,
}

\fi %end of commands specific to pdftex

%% file: artigo-igor.bbl
\begin{thebibliography}{10}
\providecommand{\url}[1]{#1}
\csname url@samestyle\endcsname
\providecommand{\newblock}{\relax}
\providecommand{\bibinfo}[2]{#2}
\providecommand{\BIBentrySTDinterwordspacing}{\spaceskip=0pt\relax}
\providecommand{\BIBentryALTinterwordstretchfactor}{4}
\providecommand{\BIBentryALTinterwordspacing}{\spaceskip=\fontdimen2\font plus
\BIBentryALTinterwordstretchfactor\fontdimen3\font minus
  \fontdimen4\font\relax}
\providecommand{\BIBforeignlanguage}[2]{{%
\expandafter\ifx\csname l@#1\endcsname\relax
\typeout{** WARNING: IEEEtran.bst: No hyphenation pattern has been}%
\typeout{** loaded for the language `#1'. Using the pattern for}%
\typeout{** the default language instead.}%
\else
\language=\csname l@#1\endcsname
\fi
#2}}
\providecommand{\BIBdecl}{\relax}
\BIBdecl

\bibitem{dahlmanet2011lte}
E.~Dahlmanet \emph{et~al.}, \emph{4G LTE/LTE-Advanced for Mobile
  Broadband}.\hskip 1em plus 0.5em minus 0.4em\relax Academic Press, UK, 2011.

\bibitem{luo2018cran}
Y.~Luo \emph{et~al.}, ``A computation workload characteristic study of c-ran,''
  \emph{International Conference on Distributed Computing Systems (ICDCS)}, pp.
  1599--1603, 2018.

\bibitem{yeoh2016oai}
C.~Y. Yeoh \emph{et~al.}, ``Performance study of lte experimental testbed using
  openairinterface,'' \emph{Performance study of LTE experimental testbed using
  OpenAirInterface}, pp. 617--622, 2016.

\bibitem{nahum2017emulation}
C.~Nahum \emph{et~al.}, ``Emulation of {4G/5G} network using
  openairinterface,'' \emph{SBrT XXXV Simp{\'o}sio Brasileiro de
  Telecomunica\c{c}{\~o}es e Processamento de Sinais}, 2017.

\bibitem{checko2015cranclassic}
A.~Checko \emph{et~al.}, ``Cloud ran for mobile networks a technology
  overview,'' \emph{International Conference on Advanced Communication
  Technology (ICACT)}, pp. 405--426, 2015.

\bibitem{docker}
\BIBentryALTinterwordspacing
``\BIBforeignlanguage{en-US}{Docker}.'' [Online]. Available:
  \url{https://www.docker.com/}
\BIBentrySTDinterwordspacing

\bibitem{dua2014virtualization}
R.~Dua \emph{et~al.}, ``Virtualization vs containerization to support paas,''
  \emph{IEEE International Conference on Cloud Engineering}, pp. 610--614,
  2014.

\bibitem{oai}
\BIBentryALTinterwordspacing
``\BIBforeignlanguage{en-US}{{The OpenAirInterface Platform} - how to use oai
  to setup c-ran}.'' [Online]. Available:
  \url{https://gitlab.eurecom.fr/oai/openairinterface5g/wikis/how-to-connect-cots-ue-to-oai-enb-via-ngfi-rru}
\BIBentrySTDinterwordspacing

\bibitem{kumar2017docker-compose}
S.~K. Pentyala, ``Emergency communication system with docker containers, osm
  and rsync,'' \emph{International Conference On Smart Technologies For Smart
  Nation (SmartTechCon)}, pp. 1064--1069, 2017.

\bibitem{chang2018hostmode}
H.~Chang \emph{et~al.}, ``Performance evaluation of open5gcore over kvm and
  docker by using open5gmtc,'' \emph{IEEE/IFIP Network Operations and
  Management Symposium}, pp. 1--6, 2018.

\bibitem{combe2016privilege}
A.~M. T.~Combe and R.~D. Pietro, ``To docker or not to docker: A security
  perspective,'' \emph{IEEE Cloud Computing}, vol.~3, no.~5, pp. 54--62, 2016.

\end{thebibliography}
